# Aurorae in Australian Aboriginal Traditions

Duane W. Hamacher

Nura Gili Centre for Indigenous Programs, University of New South Wales, Sydney, NSW, 2052, Australia

*Corresponding email: d.hamacher@unsw.edu.au*

**Abstract**

Transient celestial phenomena feature prominently in the astronomical knowledge and traditions of Aboriginal Australians. In this paper, I collect accounts of the Aurora Australis from the literature regarding Aboriginal culture. Using previous studies of meteors, eclipses, and comets in Aboriginal traditions, I anticipate that the physical properties of aurora, such as their generally red colour as seen from southern Australia, will be associated with fire, death, blood, and evil spirits. The survey reveals this to be the case and also explores historical auroral events in Aboriginal cultures, aurorae in rock art, and briefly compares Aboriginal auroral traditions with other global indigenous groups, including the Maori of New Zealand.

## 1  Introduction

Oral tradition and material culture are the mechanisms by which knowledge about the natural world is transmitted to successive generations of Aboriginal Australians. These traditions contain a significant astronomical component (Haynes 1992; Johnson 1998; Norris & Hamacher 2009) useful for navigation, calendars, and food economics (Clarke 2007; Hamacher & Norris 2011c; Hamacher 2012:71-86). Aboriginal people would move from place to place within their country to seek out food sources and shelter throughout the year. The motions and positions of celestial bodies with respect to the surrounding landscape were of great importance to this end, signaling the changing seasons and the availability of particular food sources. For example, the heliacal rising of the star Fomalhaut signals the coming of the autumn rains to the Kaurna people of the Adelaide plains (Hamacher 2012:79-82). Another example is the acronychal rising of the celestial emu (traced out by the dust lanes of the Milky Way between Crux and Sagittarius) signaling the start of the emu breeding season, when emu eggs are used as a food source (Norris & Norris 2009; Hamacher 2012:71-75). Astronomical traditions also contain a social component (Johnson 1998). Celestial objects serve as mnemonic devices for remembering laws and customs and inform traditions regarding marriage classes, totems, and ceremonies (Clarke 2007; Hamacher 2012; Fuller et al. 2013).

The study of the astronomical knowledge and traditions of Aboriginal and Torres Strait Islander people - part of the interdiscipline of *cultural astronomy* - is a growing subject of public and academic interest in Australia. One component of ongoing research in Australian cultural astronomy involves the investigation of transient celestial phenomena. This includes studies of variable stars (Hamacher & Frew 2010), eclipses (Hamacher & Norris 2011a), comets (Hamacher & Norris 2011b), meteors (Hamacher & Norris 2010) and cosmic impacts (Hamacher & Norris 2009) in Aboriginal traditions.

The main goal of this research paper is to investigate the knowledge and traditions relating to the Aurora Australis. This includes understanding how the phenomenon was and still is perceived by Aboriginal people, how knowledge of aurorae are incorporated into astronomical traditions, and how the aurora is represented in material culture (i.e. artefacts or rock art).

I begin by discussing the background theory and methodology of this study, followed by a discussion of the physics of aurorae and the solar cycle. I then discuss how previous studies of transient celestial phenomena can be combined with the physical properties of aurorae to predict how Aboriginal people might perceive this phenomenon. Next, I explore the Aboriginal traditions related to aurorae, including explanations of the phenomenon. I also investigate oral traditions for historical auroral events to determine if Aboriginal people noted or predicted the reoccurring frequency and intensity of aurorae over the 11-year solar cycle. I then search for possible representations of aurorae in rock art and discuss how Aboriginal auroral traditions compare with those of the Maori of New Zealand and other global indigenous groups that live in or near auroral zones.

## 2  Theory & Methodology

The cross-disciplinary field of cultural astronomy, including the sub-disciplines of archaeoastronomy and ethnoastronomy, is a social science informed by the physical sciences; it is a science asking social questions (Ruggles 2011). There is currently no all-encompassing theoretical framework for cultural astronomy and attempts to develop one have thus far been unsuccessful (Iwaniszewski 2011). Cultural astronomy currently relies on the methods and theories of the academic disciplines from which it draws. Generally, this includes archaeology, anthropology, and history. In this paper I draw from historic, ethnographic, linguistic, and archaeological records. I reviewed the literature for any sources containing references to aurorae (including 'southern lights' or 'sky glow') in Aboriginal cultures. These sources included ethnographies, archaeological surveys, historical documents, stories, songs, magazines, newspaper and journal articles, audio and video sources, reputable web-sources, postgraduate theses, rock art, and other artistic forms.

Since this research addresses both Western science and Aboriginal traditions regarding aurorae for a wide audience of readers, it is important to describe the phenomenon from both perspectives. For this reason, a brief, non-mathematical description of auroral physics is provided in the next section. This information is then combined with previous cultural studies of transient celestial phenomena to predict how Aboriginal people, as reported in the literature, perceive aurorae.

## 3       The Physics of Aurorae

Aurorae are light displays, generally seen at high (i.e. polar) or middle latitudes, caused when energetic charged particles in near-Earth space bombard the atmosphere, increasing the energy of oxygen and nitrogen molecules, which then emit visible light. The entire process is governed by solar activity, and in particular the solar wind, which normally is blocked from reaching Earth space by Earth's magnetic field. However, due to the configuration of the magnetic field, solar wind energy can access low altitudes near Earth's magnetic poles, where magnetosphere particles (mostly electrons and protons) can be accelerated along field lines into the upper atmosphere (> 80 km above sea level). These particles impact and energize oxygen and nitrogen molecules, which then release photons (particles of light). The photons are released in discrete packets of energy, which correspond to different colours of the electromagnetic spectrum. Oxygen emissions are generally green to dark red, depending on the amount of energy that is absorbed. Nitrogen emissions are blue to red, depending on whether the atom regains an electron after it has been ionized (blue) or whether it loses energy (red).

The colour of an aurora to an observer on the Earth's surface is dependent on the aurora's altitude. The time for oxygen ions to emit green light is only three seconds, while the time needed to emit red light is up to two minutes. Oxygen is more abundant in the upper atmosphere where collisions between molecules are low. Because of this, red colours dominate in the upper atmosphere. Green colours dominate in the lower atmosphere, where the air density is higher and collisions are more frequent.

Aurorae are referred to as *Aurorae Borealis* in the northern hemisphere and *Aurorae Australis* in the southern hemisphere. Although aurorae can technically be seen from many places on the Earth, they tend to be most active in oval ring-shaped regions located in each hemisphere about 10°-20° equator-ward from Earth's magnetic poles. The south magnetic pole is ~150 km off the coast of Antarctica towards Tasmania (64.497° S, 137.684° E) as of 2007 and the north magnetic pole is ~740 km northwest of Ellesmere Island in northern Canada as of 2012 (85.9° N, 147.0° W) (NOAA 2012). The magnetic poles are not antipodal and they wander at a relatively high rate: currently 10-15 km yr$^{-1}$ in the south and 45-62 km yr$^{-1}$ in the north (Zvera 2012).

The width and location of the auroral zones varies with the intensity of disturbance of the geomagnetic field, and the location over time of the wandering geomagnetic poles. In the northern hemisphere, the auroral ring generally covers Alaska, Canada, northern regions of the United States, northern Europe and Siberia. In the southern hemisphere, the auroral zone generally covers Antarctica and the southern fringes of Australia and New Zealand (Figure 1). Under average, magnetically undisturbed conditions observers from southern Australia and much of New Zealand only have a slight chance of witnessing an aurora (Bond & Jacka 1962; McEwan 2006).

The intensity of the solar wind and hence the rate and level of magnetic disturbance at Earth depends on the Sun's magnetic activity. This activity waxes and wanes during the solar cycle, the duration of which is approximately 11 years. Periods of high solar activity are marked by an increase in sunspots, which denote a rise in the Sun's magnetic activity. This increase in activity leads to larger and more frequent solar eruptions, such as coronal mass ejections, causing more frequent and intense magnetic storms at Earth. During solar maxima, aurorae are therefore more frequent and intense. Major coronal mass ejections can cause aurorae to be visible from areas of the Earth that rarely witness such a phenomenon. For example, a major solar event on the night of 25-26 September 1909 caused aurorae that were visible in both hemispheres, with aurorae visible as far north as Queensland (Duncan-Kemp 1952:44). The physics of solar-terrestrial magnetism and aurorae are more complex than is described in this section. The curious reader should explore Chapman (1970), Jones (1974), and Carlson & Egeland (1995) to learn more about the science of aurorae.

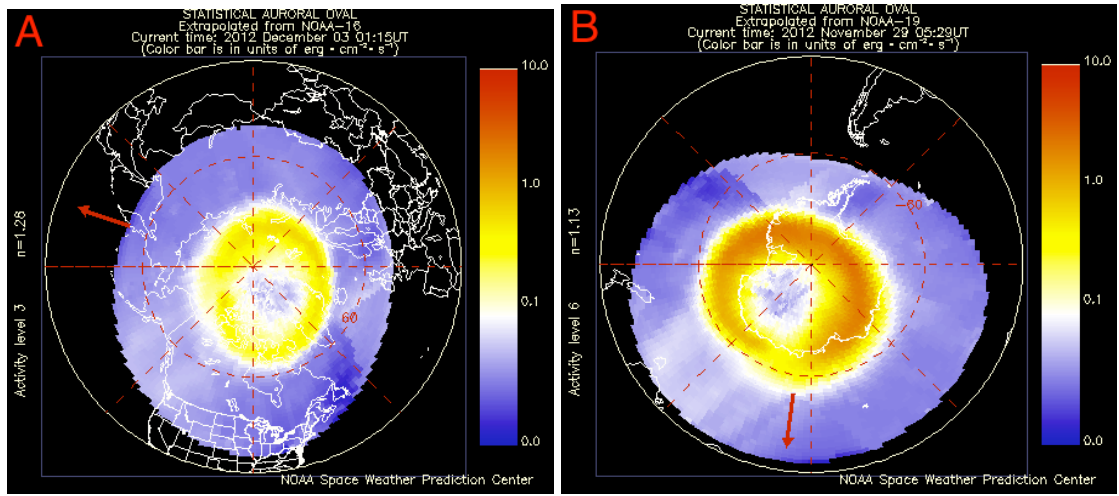

*Figure 1: (A) A snapshot of the extent and position of the auroral zone in the northern hemisphere at 23:06 UT on 2 December 2012. (B) A snapshot of the southern auroral zone at 05:29 UT on 29 November 2012. Both were relatively quiet days in terms of magnetic disturbance. Under more disturbed conditions the auroral zones are thicker and displaced equator-ward. The plots are derived from measurements obtained from the NOAA POES satellite showing the power flux of auroral activity (0 to 10 ergs cm$^{-2}$ sec$^{-1}$). Plot taken from the POES Auroral Activity page on the National Weather Service's Space Weather Prediction Centre, NOAA. URL: http://www.swpc.noaa.gov/pmap/*

## 4   Predictive Ethnoastronomy

I conducted a systematic survey of transient celestial phenomena in Australian Aboriginal cultures, with published results covering comets (Hamacher & Norris 2011a), eclipses (Hamacher & Norris, 2011b), meteors (Hamacher & Norris 2010), the Great Eruption of Eta Carinae (Hamacher & Frew 2010), and meteorite impacts (Hamacher & Norris 2009; Hamacher & Goldsmith 2013). Each of these studies revealed similarities in the ways these types of phenomena are incorporated into Aboriginal traditions. In a majority of cases, transient astronomical events were seen negatively and were often associated with evil spirits, black magic, or omens of war, death, and disease. Studies like these help cultural astronomers in the effort to develop a theoretical base for the discipline. These studies are useful for informing us on how humans think about the natural world and how they develop knowledge to explain it.

By combining these studies and taking into account the physical properties of aurorae, it may be possible to predict how people interpreted this phenomenon and which groups of people would likely have traditions about it. Six primary attributes relate aurorae to human perception:

1. The location on earth from which aurorae are visible;
2. The probability of witnessing an aurora;
3. The direction from which aurorae are visible to an observer;
4. The physical appearance of an aurora;
5. The colour of the aurora; and
6. The intensity of the aurora.

Technically, aurorae can be seen from many places on Earth, but seeing one from an area far from the auroral zone is extremely rare. Since the southern fringes of continental Australia, Tasmania, and the south island of New Zealand are near and under certain conditions within the edges of the southern auroral zone, we expect that aurorae will be incorporated into the traditions of Indigenous communities in these regions. Since aurorae are primarily seen toward the south, we expect that Indigenous knowledge of this phenomenon relates to this direction. In many Aboriginal cultures, the direction of an astronomical phenomenon is significant. For example, meteors denote the direction of an enemy in the traditions of Aboriginal people near the Tully River, Queensland (Roth, 1984: 8) and to the Ngarigo people of southeast New South Wales (Howitt, 1904: 430). To the Arrernte of the Central Desert, the tail of a comet points toward the direction of a neighboring community in which someone has died (Spencer and Gillen, 1899: 549).

The appearance of aurorae is broken into two basic categories: diffuse and discrete. As the name suggests, diffuse aurorae lack clear patterns and generally appear as a glow. They are formed when interactions between wave-particles scatter electrons parallel to the magnetic field lines. Discrete aurorae appear in various shapes, such as draperies ("curtains"), rays, or arcs. They are formed due to electrons accelerating parallel to the earth's magnetic field lines into the atmosphere. Both diffuse and discrete aurorae can appear to move, with draperies being among the most obvious and dramatic examples (see Livesey, 2001). We might expect that oral traditions describe these different types of aurorae.

Unlike the northern hemisphere, no inhabited land lies under the highest flux areas of displays visible from Australia. The aurorae visible form Australia tend to be high altitude and are generally visible lower on the horizon (i.e. not overhead). Since high altitude aurorae are dominated by oxygen-red (as discussed in the previous section), a majority of aurorae visible from Australia are reddish in colour (Figure 2). However, in low light conditions, the human eye cannot discern colour and the aurorae appear a faint white. This is different from the bright white displays that are visible under very active conditions. Aurorae can appear to be a range of colours, including red, pink, orange, green, and white. Aurorae visible from lands under the higher flux regions of the northern auroral zone tend to reflect this range of colours, particularly white and green and can be seen directly overhead. From studies of transient celestial phenomena in Aboriginal traditions, red was often associated with blood, fire, and death. For example, the red colour of the moon during a total lunar eclipse was commonly associated with blood, fire, and evil (Hamacher & Norris, 2011a). In Lardil traditions on Mornington Island, red or blue coloured meteors were associated with sickness, while white meteors were signs of good luck (Hamacher & Norris, 2010). In Aboriginal cultures near Ooldea, South Australia, the red stars Betelgeuse and Aldebaran signified fire that was cast between celestial beings engaged in battle (Hamacher, 2012:17-18). Therefore, we expect accounts of aurorae in Aboriginal traditions to relate to blood, fire, war, and death.

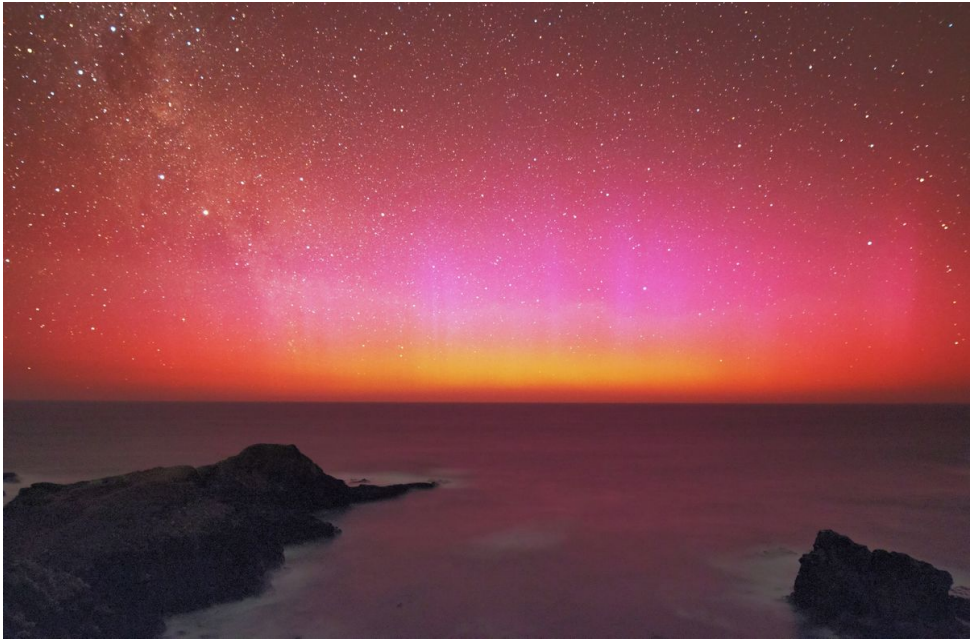

*Figure 2: The Aurorae Australis as seen from Victoria, Australia. Image courtesy of Alex Cherney.*

Aurorae also vary in visible intensity. Some are faint to the naked eye, while sometimes under very active conditions they may be bright enough to enable an observer to read a newspaper at night. We therefore expect that peoples' reactions to bright aurorae are more severe than fainter aurorae. If they relate to negative attributes, such as blood or fire, intense (bright) aurorae would probably induce a reaction of fear and panic more so than aurorae that are of low intensity. While we are confident that not all perceptions of aurorae will be negative, we expect this to be largely the case, based on previous studies of transient celestial phenomena in Aboriginal traditions (c.f. Hamacher & Norris 2009; 2010; 2011a,b).

A survey of auroral traditions from across the world (e.g. Eather 1980; Falck-Ytter, 2000; Section 9 of this paper) suggests that in areas where aurorae are a frequent occurrence, they possess benign attributes. But in areas where they are less frequent, they tend to be associated with evil, omens, and death. Since far southern Australia is at the edge of the auroral zone, we expect aurorae to be generally negative in perception since they are infrequent.

## 5      Aurorae in Aboriginal Astronomical Traditions

I analysed the sources of data mentioned in Section 2 and found they reveal major themes in the perceptions of aurorae, as predicted in Section 4. Not all views were associated with negative attributes, but most were associated with blood, fire, and death. Aurorae cause great fear in the people who witness them and in some communities they are taboo – only to be viewed and interpreted by initiated elders. Interpretations of auroral displays vary widely, even within the same communities.

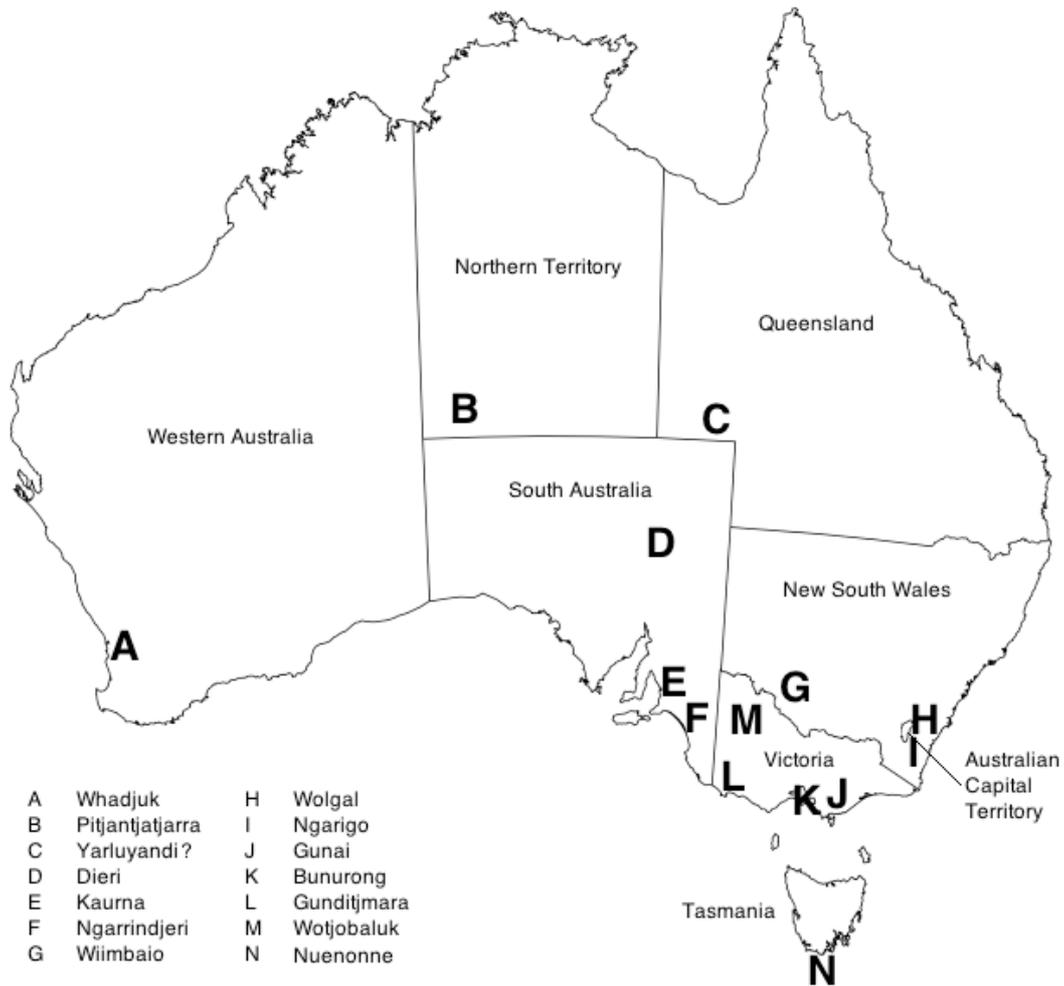

*Figure 3: Places in Australia from which Aboriginal traditions describing aurorae were found. Accounts of aurora are found in all states and territories, although they are restricted to the southern half of the continent, consistent with the rarity of these events at lower latitudes.*

Stories or accounts of aurorae were found in all Australian states and territories (Figure 3), but most of the accounts and oral traditions are from Aboriginal groups in Victoria and South Australia. All accounts were from regions south of the Tropic of Capricorn. The data are broken down by theme in Subsections 5.1-5.3. This study produced 26 literary sources representing 13 identified Aboriginal groups and five unidentified groups. The identified groups are the Gunditjmara, Gunai, Bunurong, and Wotjobaluk of Victoria, the Dieri, Narringeri, and Kaurna of South Australia, the Wiimbaio, Wolgal, and the Ngarigo of New South Wales, the Pitjantjatjarra of the Northern Territory, the Nuenonne of Tasmania, and the Whadjuk of Western Australia. Of the five unidentified groups, three were from Tasmania and one each from Queensland and New South Wales. The Queensland community is probably Yarluyandi, but this is uncertain.

Some of the accounts of aurorae were vague or did not attribute either positively or negatively to aurorae. For example, the Nuenonne people of Bruny Island, Tasmania call the Aurora *nummergen* (Wilson 1999), but no associated stories or interpretations of this name are provided. A brilliant, multi-coloured aurora visible from Hobart, Tasmania on 4 September 1851 made a crackling sound that was described by the local Aboriginal people as similar to snapping their fingers (Anon 1877).

### 5.1 Fire, Smoke & Ash

Since aurorae generally appear red in the sky as seen from Australia, they are commonly associated with fire. This association may include flames, smoke, or ashes. For example, the Gunditjmara people of coastal western Victoria called aurorae "Puae buae", meaning "ashes" (Dawson, 1881:101). The Gunai of Gippsland, eastern Victoria perceived aurorae as bushfires in the spirit world (Massola, 1965:213). They also attributed it to the fire of an ancestral hero warning of a coming catastrophe (Worms, 1986:112). When Dieri people of Cooper's Creek, South Australia saw an auroral display in 1869, they claimed it was a Kootchee (an evil spirit) creating a large fire (Smyth, 1878:458). Similarly, Narringeri people of Encounter Bay, South Australia interpreted an aurora seen over Kangaroo Island (Karta) as the campfires of spirits in the "Land of the Dead" located in the heavens (Tindale, 1974). The Narringeri considered Kangaroo Island to be the land of the dead (Tindale & Maegraith, 1931; Berndt, 1940:182). The island was uninhabited when Matthew Flinders first arrived but archaeological data revealed that Aboriginal people lived on the island as long ago as 16,000 years BP but mysteriously disappeared some 2,000 years ago (Draper, 1987).

The solar eruption from 25-26 September 1909 caused bright aurorae that were visible across the northern and southern hemispheres (Silverman, 1995). On 24 September, Duncan-Kemp (1952:44) described the appearance of bright aurorae visible from Windorah in far southwest Queensland. A group of Aboriginal people (possibly the Yarluyandi or Karuwali people) said the aurorae were the "feast fires" of the *Oola Pikka* - ghostly beings who spoke to the people through auroral "flames." Only male elders were permitted to view the display and interpret their messages, as tribal law forbade women or uninitiated men from seeing the sacred lights. When the aurorae appeared in the sky, the Aboriginal women would turn away.

Several accounts from the Gunai people describe aurorae as a physical manifestation of a powerful sky deity's anger. This seems to be slightly different than the "bushfires in the spirit world" as described by Massola (1965:213). The accounts describe the deity as *Mungan Ngour.* Mungan set the rules for the initiation of boys into manhood and put his son, Tundun, in charge of the ceremonies. Information about male initiation ceremonies was restricted to men and taboo to women. A person in the community revealed initiation secrets to the women, greatly angering Mungan. In a rage, he cast down a great fire to destroy the Earth, which the people saw as an aurora.

According to Howitt (1904:430), the appearance of Mungan's Fire caused a reaction of fear, prompting the people to shout "send it away; do not let it burn us up!" As they yelled this, they swung a dead hand, called a *bret,* at the portent. The hand served as a charm and a warning device. When a relative or close friend passed away, his or her hands were removed then smoked and dried over a fire. They were then suspended by a cord of possum fur, fingers down, around the neck over the left arm. The *bret* was believed to pinch or tap the wearer whenever danger approached. The wearer would then face in different directions, holding the *bret* in front of him. When it shook violently the man knew the direction of the danger.

Massola (1968:162) explained that Mungan's fire caused everyone to turn on each other; men speared one another and killed their wives while mothers killed their children. The aurora filled the whole of the sky from the land to the sky and was followed by a tsunami that rushed over the land and drowned most of the people (Thomas 1906a:219). Those who survived became the Muk-Kurnai, the "Superior Animals." Tundun, the Great Man's son, and his wife, became porpoises. The Great Man then went to the sky. After his anger subsided, he allowed the men to again carry on the ceremonies on the condition that they not tell women initiation secrets. If his laws and customs are disregarded, he shows his anger by lighting up the sky at night with the fires of the aurora (Keen, 2004:214-215).

Thomas (1906b:143) noted that an auroral display caused the Gunai people to undergo a sense of temporary promiscuity. To avert the evil of Mungan's fire, elder men would exchange wives in a practice called *beamu* (Keen, 2004: 182). Howitt (1891:101) also noted this practice, explaining it as a reversion to the ancient custom of group marriage. This practice was also found among the Dieri people (Montagu, 2004:219). Aurora traditions among the Bunurong people of the Mornington Peninsula, southeast of Melbourne are similar to the Gunai (McCrae 1845).

A final account that attributes aurorae to fire is from the Pitjantjatjarra people near Uluru (Ayer's Rock) in the Northern Territory. The story describes hunters breaking a taboo by cooking a sacred emu (*kalaya*). They saw smoke rise to the south, towards the land of *Tjura* (a Pitjantjatjarra word meaning "glowing light visible at night"; Goddard 1992:160). This was the glow of the Aurora Australis, which the Pitjantjatjarra believed were poisonous flames (Harney 1960:74). In the story, the flames served as a portent of punishment to the hunters.

### 5.2 Blood and Death

The appearance of an unexpected transient event in the sky is often met with fear (e.g. Hamacher & Norris 2010, 2011a,b). The appearance of a red aurora in the southern evening skies as seen from the central-western coast of Tasmania caused astonishment and incited shouting by Aboriginal people (Anon 1838). The Whadjuk people in Perth had a similar reaction on night of 11 July 1838 when a bright auroral display was visible (Anon 1838). Similar to the appearance of comets and meteors, an auroral display was an omen to the Dieri people that a person in a neighboring community had condemned someone to death (Frazer *et al.* 1895:175-176). Thus, the appearance of an aurora – as with many transient phenomena - was met with great fear.

To the Dieri people, the Aurora Australis is called *pilliethillcha*. Whenever an auroral display is seen, it causes great fear and anxiety. The Dieri believe it is a warning from the devil (Kootchie) to keep a strict watch, as an armed party (*pinya*) is killing some one for breaking traditional laws. Aboriginal people in the camp then huddle together, when one or two step out and perform a ceremony to charm the Kootchie (Gason, 1879:297). Fear of an aurora was developed and utilised to control behavior and social standards. Breaking traditional laws would result in a *pinya* coming to kill the lawbreakers when they least expect.[1]

The red colour of most aurorae visible from Australia was commonly associated with blood and death, as anticipated in Section 4. Auroral displays represented blood that was shed by warriors fighting a great battle or by massacre victims rising to the sky. This view was shared by Aboriginal people in the Riverine areas of New South Wales and South Australia, the Ngarigo and Wolgal people near Canberra, the Wotjobaluk people of northwestern Victoria (Howitt 1904: 430; Gibbs 1974:50), and the Dieri people (Gason, 1879: 297).

### 5.3 Spirits and Omens

Smith (1913) records accounts from two unspecified Aboriginal groups that attribute aurorae to the spirits of ancestors dancing in the sky or a portent of trouble. Aurorae served as omens of disease to the Wiimbaio near the junction of the Murray and Darling Rivers (Thomas 1906b:143). A bright pink aurora visible from Adelaide on 7 February 1840 (Anon 1840) was believed to be the harbinger of a plague. The nearby Putpa, Wirra, and Marimeyunna clans believed aurorae were caused by sorcerers from the north and was an omen (Schurmann 1987:86). The Ngarrindjeri people of Point McLeay mission, South Australia shared a similar view. A combination of rare astronomical phenomena caused fear and anxiety to the local Ngarrindjeri people in 1859. In August of that year, an aurora was visible around the time of a lunar eclipse (Taplin 1859:2 September 1859; Anon 1859). The Ngarrindjeri believed this signaled the arrival of dangerous spirit beings they dubbed "wild blackfellows." These were Aboriginal people living outside the areas of European settlements (Merlan 1994). Like their Wiimbaio neighbors, the Ngarrindjeri considered "wild blackfellows" to be great sorcerers with supernatural powers and were greatly feared.

### 6 Historical Auroral Events

Aurorae and sunspot cycles have been recorded throughout human history, particularly in eastern Asia (Keimatsu 1970-1976; Lee et al., 2006; Matsushita, 1956; Stephenson and Willis, 1999, 2008; Willis and Davis, 2013; Willis and Stephenson, 2000, 2001; Willis *et al.* 2005, 2007). Since there are no written records from Australia prior to colonisation, it is difficult to connect historical aurorae in antiquity with Aboriginal traditions since the latter do not necessary record a single point in linear time. It is probable that aurorae visible in antiquity were the basis, or re-emphasis, of auroral traditions in Australia. However, since colonisation of Australia by the British in 1788, historical records of auroral events and written accounts of Aboriginal traditions can be matched.

These historical records revealed six auroral events noted by Aboriginal people. They occurred during major solar eruptions in 1838, 1840, 1851, 1859, 1869, and 1909. Each of these years coincides with a peak in the solar cycle (Figure 4). A bright auroral display on 11 July 1838 was visible from Tasmania to Perth (Anon 1838) and a bright pink aurora was visible from Adelaide on 7 February 1840 (Anon 1840). An aurora seen on 4 September 1851 from Hobart made sounds the local Aboriginal people described as being similar to snapping fingers. The aurora visible on 2 September 1859 from the Point McLeay mission, South Australia (Taplin 1859: 4-7 June) coincided with the strongest geomagnetic storm ever recorded (dubbed the "1859 solar super-storm" or "Carrington Event"). The bright aurorae witnessed by the Dieri people in 1869 (Smyth 1878:458) were part of the many auroral displays visible across the globe in April/May and September 1869 (Anon 1869; Prowde et al. 1869; Murray 1869; Tebbutt 1870). Major solar eruptions peaking on 25-26 September 1909 produced bright, multi-coloured aurorae visible as far north as southern Queensland (Duncan-Kemp 1952:44). It is worth noting that all three auroral displays occurred during the month of September. This accords with the general increase in geomagnetic disturbance near equinoxes (see below). Since the auroral zone only barely covers the southern edge of Australia, the discovery that historical literature only records accounts of aurorae in Aboriginal traditions during peaks in the solar cycle is expected.

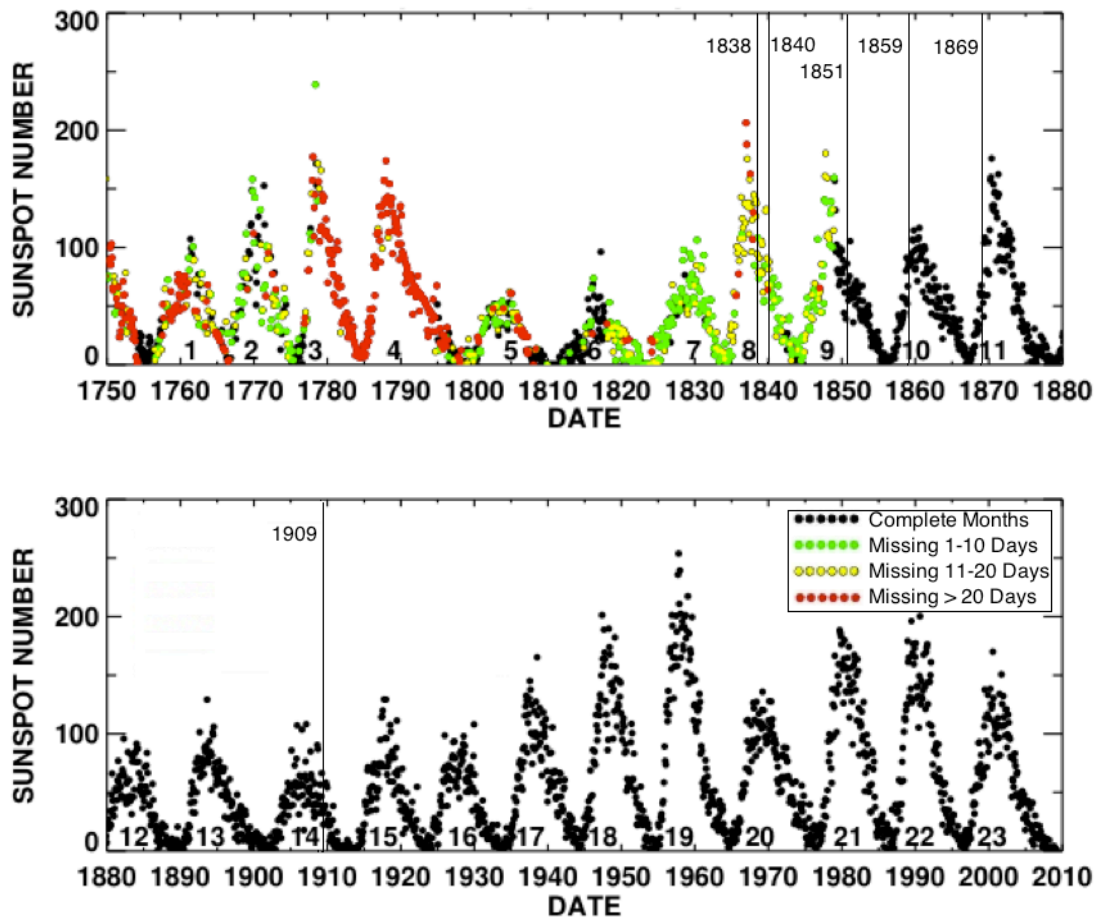

*Figure 4: The number of sunspots observed between 1750 and 2010. After Hathaway (2010: Figure 2). The vertical lines show the years when historic aurorae were described in Aboriginal culture from Section 6.*

It should be noted that a significant geomagnetic storm led to strong aurorae visible around the world on 4-5 February 1872 (Silverman 2008). Accounts from Australia describe brilliant aurorae in Perth, Adelaide, and Melbourne on these days. While the aurorae may have been visible at more northerly latitudes (*ibid*), no accounts were recorded from more northerly latitudes in the TROVE newspaper database. There were no recorded accounts of Aboriginal perspectives of the aurora, although it is almost certain that Aboriginal people would have seen it in the skies.

## 7    Aboriginal recognition of the 11-year solar cycle?

Aurorae are connected to the solar sunspot cycle and are more frequent and intense during the solar maxima, which occur every 11 years. Since Aboriginal people were careful observers of natural phenomena, we might expect to find a description of this cycle in Aboriginal traditions.

A survey of the literature revealed only one case of Aboriginal people recording the solar cycle in their traditions. According to Bodkin (2008:70) the Dharawal people south of Sydney used the appearance of the aurora to announce the start of the 11-12 year Mudong weather cycle. The aurorae appear during the annual season of Ngoonungi. Ngoonungi is the period of gradual warmth, during September and October. The appearance of aurorae signalled the start of the first of the eight Mudong cycle phases: *Gadalung Burara* – the hot and dry phase. This season can last up to 20 complete lunar cycles ("20 moons") but no more (Bodkin 2008:72). In an interview with CNN in 2003[2], Bodkin claimed that the 11-year Mudong cycle started in 2001 with the appearance of aurorae. This coincided with the peak in sunspots and the start of the last solar cycle (Figure 4). Bodkin stated that the aurorae are seen less frequently because of increasing light pollution.

In some media interviews following the publication of her book on Dharawal seasons and climatic cycles[3], Bodkin claimed that during October aurorae were visible in the western skies as well as to the south. Historical records describe aurorae visible to the west from the Hunter Valley north of Sydney (Anon 1846). It is worth noting that geomagnetic storms tend to peak in the months around the equinoxes, in particular March-April and September-October (Stamper et al. 1999; Papitashvili et al. 2000). This is when both hemispheres of Earth are most uniformly exposed to the solar wind and its embedded interplanetary magnetic field. Accordingly, many of the aurorae identified in Aboriginal traditions were linked to auroral displays during, or around, these months, particularly September.

A higher than normal rate of observed aurorae, especially those seen from the Sydney region, could be useful for recognising the 11-year solar cycle. There has been much discussion on whether climate follows an 11-12 year cycle in phase with the solar cycle (e.g. Haigh 2007, Weart 2013). Such an effect has not been established and is a topic of current research.

## 8    Representations of Aurorae in Rock Art

Clear representations of aurorae in Australian rock art have not been reported in the literature. Plausible, but unconfirmed, representations of aurorae in Australian rock art are reported from a group of researchers led by Anthony Peratt (Peratt, 2003; Peratt et al. 2007; Van Der Sluijs & Peratt, 2010). Peratt and his colleagues claim that in antiquity, intense solar activity directed a significant amount of plasma energy towards the Earth, interacting with Earth's magnetic field and causing brilliant aurorae that would have been seen worldwide. Based on the visible shapes derived from computer simulations of these proposed auroral events, Peratt and his colleagues then compare the auroral shapes to rock art motifs from around the world. The team notes several motifs that are evident in rock art around the world and suggest that they depict this auroral display (Figure 5).

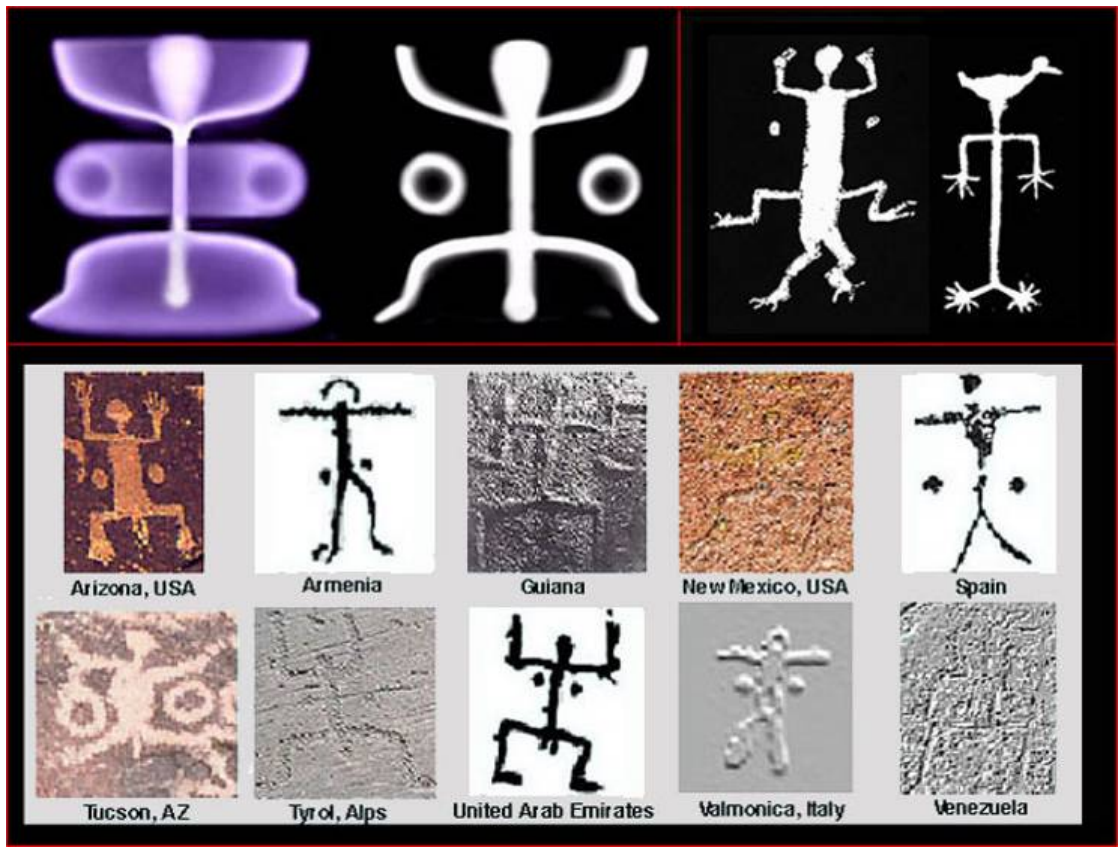

*Figure 5: A computer simulation of one of the shapes caused by the proposed high-intensity auroral display (top left), an interpretation of what it may have looked like (top middle), and various rock art motifs that Peratt and his colleagues believe may represent this phenomenon (top right, bottom). Similar motifs are found in Australia (e.g. McCarthy 1976), although any connection to the proposed auroral display is speculative. After Peratt (2003).*

The interested reader is directed to the papers of Peratt and colleagues for an in-depth explanation and analysis of this work. At this stage there is no clear evidence that any rock art motifs in Australia represent these auroral displays or that the artistic motifs described by Peratt and colleagues represent these proposed auroral events. Without oral traditions or ethnohistoric evidence to explain the meaning of artistic forms, these motifs are open to interpretation. It is worth noting that auroral displays are affected by increases and decreases in the Earth's magnetic field. The magnetic poles reverse on long timescales (450,000 years on average), with the magnetic field decreasing significantly before reversing. This could result in significant auroral displays. The last major geomagnetic excursion occurred 780,000 years BP, but recent evidence indicates that an excursion occurred approximately 41,000 years BP (Nowaczyk *et al*. 2012). This is within the period of human habitation of Australia, although we urge caution in stretching these claims too far.

## 9      Other Cultural Views of Aurorae

Aurorae are included in the oral traditions and mythology of numerous cultures across the world (see Holzworth 1974; Stothers 1979; Akasofu 1979). To the Inuit of the Hudson Strait in the far north of Quebec, Canada, aurorae represented the torches of spirits who were leading the souls of the recently deceased to paradise (Holzworth 1974). The Inuit of the Coronation Gulf in eastern Nunavut, Canada perceived aurorae as a manifestation of the spirits that brought good weather (Weyer 1969:243). In Point Barrow, Alaska, aurorae were greatly feared and the people armed themselves for protection (*ibid*). Associations with death, spirits, or battles were found among the people of Scotland (Mackenzie 1935:222), Siberia (MacCulloch 1964:398), Finland (*ibid*:81), and Estonia (*ibid*:81).

The Maori of New Zealand call aurorae *Tahunui-a-rangi*, which roughly translates to "great glowing sky" (Reed 1999:194-195; Moorfield 2011). Descriptions of aurorae in Maroi traditions are numerous. For example, a Maori man from Whanganui on the southern coast of the North Island was interviewed in 1869 and explained that when their ancestors migrated to New Zealand, one waka (canoe) continued sailing south, settling a land in the far south (Best 1955:71). The Maori see aurorae as a reflection of the fires cast by these ancestors, which signal their presence. Another story describes aurorae as Maru, the name of a Maori war-deity (Craig, 1989:160). According to Kingsley-Smith (1967), *"anyone going on the war-path and seeing an aurora would not continue on his way, for such conduct would have been considered suicidal."* A Maori person serving as a medium of the god Maru was shot by Lieutenant Colonel John H.H. St. John (1836-1876) at Ruatahuna on the North Island in 1869.

## 10     Discussion & Conclusion

Accounts and descriptions of aurorae are found in the oral traditions of Aboriginal Australians across the southern half of the continent. Significant solar eruptions caused auroral displays that were visible as far north as southern Queensland, but no accounts are taken from areas north of the Tropic of Capricorn. As predicted in Section 4, most Aboriginal accounts describe aurorae in negative terms and associate them with blood, death, fire, or evil spirits. Aurorae are also associated with a southerly direction. This is due largely to their generally reddish appearance on the southern horizon. A comparison of cultural interpretations of aurorae from other parts of the world, particularly New Zealand, showed a similar association.

Aurorae play an important role in the oral traditions of Aboriginal Australians. Aurora traditions provide us with a more complete understanding of Aboriginal sky knowledge and cultural astronomy. Aboriginal views of the phenomenon are similar to the Maori of New Zealand and other comparable cultural groups across the world.

**Acknowledgements**

I would like to thank Fred Menk, David Willis, Wayne Orchiston, and the referees for their helpful comments. This research made use of the following databases: TROVE (National Library of Australia), JSTOR, Mura (Australian Institute for Aboriginal and Torres Strait Islander Studies), and ADS – the Astrophysics Database System (Harvard-Smithsonian Centre for Astrophysics).

**Notes**

1. An account of retribution from a sky deity for careless behavior is reported by Berndt (1947). A Wiradjuri man (central New South Wales) claimed that when someone is careless, Kurikuta would come down from the sky, turning the night into day. Berndt suggests this is a reference to an aurora. Kurikuta, an emu, was the wife of the sky deity Baiame. She had a quartz body and when she came down from the sky, she was seen as a brilliant flash of light that made a great thunderous sound (p. 28). Given her association with a brilliant flash of light accompanied by the sound, it is probably a reference to lightning or possibly a fireball/airburst (bright/exploding meteor) as opposed to an aurora. Quartz was believed to be a material manifestation of celestial deities that were brought to Earth by meteors (Hamacher & Norris 2010).
2. http://edition.cnn.com/2003/TECH/science/03/18/offbeat.weather.aborigines.reut/
3. http://www.murrindindiclimatenetwork.org.au/24/Fran-Bodkin/

**References**

Akasofu, S., 1979. Aurora Borealis: the Amazing Northern Lights. *Alaska Geographic* 6(2), 1-95.

Anonymous, 1838. Perth Gazette (Perth, Western Australia: published from 1848 – 1864). Wednesday, 11 July 1838, p. 3.

Anonymous, 1840. Launceston Advertiser (Launceston, Tasmania: published from 1829–1846), Thursday 13 February 1840, page 3.

Anonymous, 1846. The Maitland Mercury (Maitland, NSW: published from 1843–Present), Wednesday 9 September 1846, p. 2

Anonymous, 1859. Aurora Australis. *The Argus* (Melbourne, Victoria: published from 1848–1956), Thursday 1 September 1859, p. 5.

Anonymous, 1869. Aurora Australis. *South Australian Register* (Adelaide, SA, published from 1839 – 1900), p. 3.

Anonymous, 1877. The Drought. *The Brisbane Courier* (Brisbane, Qld, published from 1864–1933), Saturday 8 September 1877, p. 3.

Berndt, R.M., 1940. Some Aspects of Jaralde Culture, South Australia. *Oceania*, 11(2), 164-185.

Berndt, R.M., 1947. Wuradjeri Magic and "Clever Men" (Continued). *Oceania*, 18(1), 60-86.

Bodkin, B., 2008. *D'harawal: seasons and climatic cycles*. Sydney: F. Bodkin & L. Robertson

Bond, F.R., and Jacka, F., 1962. Distribution of Auroras in the Southern Hemisphere. II. Nightly Probability of Overhead Aurora. *Australian Journal of Physics*, 15, 261-272.

Carlson, H.C. Jr., and Egeland, A., 1995. The Aurora and the Auroral Ionosphere. In M.G. Kivelson and C.T. Russell (Eds) *Introduction to Space Physics*. Cambridge University Press. Pp. 459-503.

Chapman, S., 1970. Auroral Physics. *Annual Review of Astronomy and Astrophysics,* 8, 61-86.

Clarke, P.A., 2007. An Overview of Australian Aboriginal Ethnoastronomy. *Archaeoastronomy*, 21, 39-58.

Craig, R.D., 1989. *Dictionary of Polynesian Mythology*. New York: Greenwood Press.

Draper, N., 1987. Context for the Kartan: A Preliminary Report on Excavations at Cape du Couedic Rockshelter, Kangaroo Island. *Archaeology in Oceania*, 22(1), 1-8.

Duncan-Kemp, A., 1952. *Where Strange Paths Go Down*. Sydney: W.R. Smith and Paterson.

Gason, S., 1879. The manners and customs of the Dieyerie tribe of Australian Aborigines. In Woods, J.D. (edt) *The Native Tribes of South Australia*. Adelaide: E.S. Wigg & Son, pp. 253-257.

Goddard, C., 1992. *Pitjantjatjara/Yankunytjathara to English Dictionary*. Alice Springs: Institute for Aboriginal Development.

Haigh, J.D., 2007. The Sun and the Earth's Climate. *Living Reviews in Solar Physics,* 4, 1-64. URL: http://www.livingreviews.org/lrsp- 2007- 2

Hamacher, D.W., 2012. *On the Astronomical Knowledge and Traditions of Aboriginal Australians*. Doctor of Philosophy Thesis (by publication), Department of Indigenous Studies, Macquarie University, Sydney.

Hamacher, D.W., and Frew, D.J., 2010. An Aboriginal Australian record of the Great Eruption of Eta Carinae. *Journal of Astronomical History & Heritage*, 13(3), 220-234.

Hamacher, D.W., and Norris, R.P., 2011a. Eclipses in Australian Aboriginal Astronomy. *Journal of Astronomical History & Heritage*, 14(2), 103-114.

Hamacher, D.W., and Norris, R.P., 2011b. Comets in Australian Aboriginal Astronomy. *Journal of Astronomical History & Heritage*, 14(1), 31-40.

Hamacher, D.W., and Norris, R.P., 2011c. "Bridging the Gap" through Australian Cultural Astronomy. In Ruggles, C.L.N. (edt) *Archaeoastronomy & Ethnoastronomy:*

*building bridges between cultures*. Cambridge University Press. Pp. 282-290.

Hamacher, D.W., and Norris, R.P., 2010. Meteors in Australian Aboriginal Dreamings. *WGN - Journal of the International Meteor Organization*, 38(3), 87-98.

Hamacher, D.W., and Norris, R.P., 2009. Australian Aboriginal Geomythology: eyewitness accounts of cosmic impacts? *Archaeoastronomy*, 22, 60-93.

Harney, W., 1960. Ritual behaviour at Ayers Rock. *Oceania*, 31(1), 63-76

Haynes, R.D., 1992. Aboriginal astronomy. *Australian Journal of Astronomy*, 4, 127-140.

Hathaway, D.H., 2010. The Solar Cycle. *Living Reviews of Solar Physics*, 7, 1-65.

Holzworth, R., 1974. Folklore and the aurora. *History of Geophysics*, 1, 41-43.

Iwaniszewski, S., 2011. The sky as a social field. In Ruggles, C.L.N. (edt) *Archaeoastronomy & Ethnoastronomy: building bridges between cultures*. Cambridge University Press. Pp. 30-37.

Lee, E.H., Ahn, Y.S., Yang, H.J., and Chen, K.Y., 2004. The sunspot and auroral activity cycle derived from Korean historical records of the 11th - 18th century. *Solar Physics*, 224, 373-386.

Johnson, D., 1998. *The Night Skies of Aboriginal Australia: A Noctuary*. Oceania Monograph #47. Sydney: University of Sydney Press.

Jones, A.V., 1974. *Aurora*. Dordrecht: Reidel.

Keimatsu, M., 1970-1976. A chronology of aurorae and sunspots observed in China, Korea and Japan. *Annals of Science, Kanazawa University*. Part I: 7, 1-10 (1970); Part II: 8, 1-16 (1971); Part III: 9, 1-36 (1972); Part IV: 10, 1-32 (1973); Part V: 11, 1-36 (1974); Part VI: 12, 1-40 (1975); Part VII: 13, 1-32 (1976).

Kingsley-Smith, C., 1967. Astronomers in puipuis. Maori Star lore. Southern Stars, 22, 5-10.

Livesey, J.K., 2001. *Aurorae Section: observing the visible aurora*. British Astronomical Association, URL: www.britastro.org/aurora/ accessed 5 July 2013.

MacCulloch, C.J.H. (Edt), 1964. *Mythology of All Races, Vol. 4*. New York: Cooper Square.

Mackenzie, D.A., 1935. *Scottish Folklore and Folklife*. London: Blackie & Son.

Massola, A., 1968. *Bunjil's Cave*. Melbourne: Lansdowne Press.

Massola, A., 1965. Some superstitions current amongst the Aborigines of Lake Tyers.

*Mankind*, 6(5), 211-214.

Matsushita, S., 1956. Ancient aurorae seen in Japan. *Journal of Geophysical Research*, 61, 297-302.

McCarthy, F.D., 1976. *Rock art of the Cobar Pediplain in central western New South Wales*. Canberra: Australian Institute for Aboriginal Studies.

McEwan, D.J., 2006. *Aurora*. In Riffenburgh, B. (edt) *Encyclopedia of the Antarctic*. Oxford: Routledge.

Merlan, F., 1994. Narratives of survival in the post-colonial North (Aboriginal Histories, Aboriginal Myths). *Oceania*, 65(2), 151-174

Montagu, A., 2004. *Coming Into Being Among The Australian Aborigines: The Procreative Beliefs Of The Australian Aborigines*. Oxford: Routledge.

Moorfield, J.C., 2011. *Te Aka Māori-English, English-Māori Dictionary and Index*. Pearson: New Zealand.

Murray, T., 1869. Accounts of Aurorae Australis. *Sydney Morning Herald,* Tuesday, 10 August 1869, p. 3.

NOAA, 2012. *Geomagnetism Frequently Asked Questions*. National Geophysical Data Centre, National Oceanic & Atmospheric Administration (NOAA), Washington, D.C. URL: http://www.ngdc.noaa.gov/geomag/faqgeom.shtml, accessed on 29 November 2012.

Norris, R.P., and Hamacher, D.W., 2009. The Astronomy of Aboriginal Australia. In Valls-Gabaud, D., and Boksenberg, A. (eds.) *The Role of Astronomy in Society and Culture*. Cambridge University Press. Pp. 39-47.

Nowaczyk, N.R., Arz, H.W., Frank, U., Kind, J., and Plessen, B., 2012. Dynamics of the Laschamp geomagnetic excursion from Black Sea sediments. *Earth and Planetary Science Letters*, 54, 351-352.

Papitashvili, V.O., Papitashva, N.E., and King, J.H., (2000). Solar cycle effects in planetary geomagnetic activity: Analysis of 36-year long OMNI dataset. *Geophysical Research Letters*, 27(17), 2797–2800.

Peratt, A. 2003. Characteristics for the Occurrence of a High-Current, Z-Pinch Aurora as Recorded in Antiquity. *Institute of Electrical and Electronics Engineers Transactions on Plasma Science*, 31(6), 1192-1214.

Peratt, A. et al 2007. Characteristics for the Occurrence of a High-Current Z-Pinch Aurora as Recorded in Antiquity Part II - Directionality and Source. *Institute of Electrical and Electronics Engineers Transactions on Plasma Science*, 35(4), 778-807.

Prowde, R., Johnson, S., Reside, J., and Lawton, W., 1869. Correspondence - Grand Auroral Display, May 13, 1869. *Astronomical Register*, 7, 135-138.

Ruggles, C.L.N., 2011. Pushing back the frontiers or still running around the same circles? 'Interpretative archaeoastronomy' thirty years on. In Ruggles, C.L.N. (edt) *Archaeoastronomy & Ethnoastronomy: building bridges between cultures*. Cambridge University Press. Pp. 1-18.

Schurmann, E.A., 1987. *I'd rather dig potatoes*. Adelaide: Lutheran Publishing House.

Silverman, S.M., 1995. Low latitude auroras: the storm of 25 September 1909. *Journal of Atmospheric and Terrestrial Physics*, 57(6), 673-685.

Silverman, S.M., 2008. Low-latitude auroras: The great aurora of 4 February 1872. *Journal of Atmospheric and Solar–Terrestrial Physics*, 70(10), 1301-1308.

Stamper, J., Lockwood, M. and Wild, M.N., 1999. Solar causes of the long-term increase in geomagnetic activity. *Journal of Geophysical Research*, 104(A12): 28, 325–28, 342

Stephenson, F.R., and Willis, D.M., 1999. The earliest drawing of sunspots. *Astronomy & Geophysics*, 40, 6.21-6.22.

Stephenson, F.R., and Willis, D.M., 2008. "Vapours like fire light" are Korean aurorae. *Astronomy & Geophysics*, 49, 3.34-3.38.

Stothers, R., 1979. Ancient Aurorae. *Isis*, 70(1), 85-95.

Taplin, G., 1859. Taplin Journals 1859-1879: 4-7 June 1859 and 2 September 1859. Adelaide: Mortlock Library.

Tebbutt, J., 1870. Aurora Australis. *Sydney Morning Herald*, Monday, 31 October 1870, p. 2.

Thomas, N., 1906a. The native races of the British Empire - Natives of Australia, London: Archibald Constable

Thomas, N., 1906b. *Kinship organisations and group marriage in Australia*. Cambridge University Press.

Tindale, N.B., 1974. Journal entry for 2 July 1974. In Canberra and California Journal, Vol. 1, 1974–1975. Adelaide: South Australian Museum Archives.

Tindale, N.B., and Maegraith, B.G., 1931. Traces of an extinct Aboriginal population on Kangaroo Island. *Records of the South Australian Museum*, 4, 275-289.

Van Der Sluijs, M.A., and Peratt, A.L., 2010. Searching for Rock Art Evidence for an Ancient Super Aurora. *Expedition*, 52(2), 33-42.

Weart, S., 2013. Changing Sun, Changing Climate? *The Discovery of Global Warming* website, American Institute of Physics. URL: http://www.aip.org/history/climate/solar.htm#N_1_

Weyer, E.M., 1969. *The Eskimos, Their Environment and Folkways.* Hamden, CN: Archon.

Willis, D.M., and Davis, C.J., 2013. Evidence for recurrent auroral activity in the twelfth and seventeenth centuries. In Orchiston W., Green, D., and Strom. R. (eds.). *StephensonFest: Studies in Historical Astronomy*. New York: Springer, in press.

Willis, D.M., and Stephenson, F.R., 2000. Simultaneous auroral observations described in the historical records of China, Japan and Korea from ancient times to AD 1700. *Annales Geophysicae*, 18, 1-10.

Willis, D.M., and Stephenson, F.R., 2001. Solar and auroral evidence for an intense recurrent geomagnetic storm during December in AD 1128. *Annales Geophysicae*, 19, 289-302.

Willis, D.M., Stephenson, F.R., and Fang, H., 2007. Sporadic aurorae observed in East Asia. *Annales Geophysicae*, 25, 417-436.

Willis, D.M., Armstrong, G.M., Ault, C.E., and Stephenson, F.R., 2005. Identification of possible intense historical geomagnetic storms using combined sunspot and auroral observations from East Asia. *Annales Geophysicae*, 23, 945-971.

Zvera, T., 2012. Motion of North and South magnetic poles in 2001-2009. European Geosciences Union General Assembly 2012, held 22-27 April 2012 in Vienna, Austria, p. 11236.

**Author Biography**

Dr. Duane Hamacher is a Lecturer in the Nura Gili Indigenous Centre at the University of New South Wales in Sydney, Australia. His research and teaching focuses on the astronomical knowledge and traditions of Indigenous Australians. He works as a consultant curator and astronomy educator at Sydney Observatory, and is the founder and Chair of the Australian Society for Indigenous Astronomy.